\documentclass[11pt,sort&compress]{elsarticle}
\usepackage[paper=letterpaper,top=24mm, bottom=26mm, left=26mm, right=26mm]{geometry}
\makeatletter
\def\ps@pprintTitle{%
 \let\@oddhead\@empty
 \let\@evenhead\@empty
 \def\@oddfoot{\centerline{\thepage}}%
 \let\@evenfoot\@oddfoot}
\makeatother
\usepackage{amsmath}
\usepackage{amsfonts}
\usepackage{slashed}
\usepackage{accents}
\usepackage{amssymb}
\usepackage{mathrsfs}
\usepackage{mathtools}
\usepackage[LGR,T1]{fontenc}
\usepackage[latin1]{inputenc}
\let\oldbibliography\thebibliography
\renewcommand{\thebibliography}[1]{%
  \oldbibliography{#1}%
  \setlength{\itemsep}{1.4pt}%
}
\usepackage[cal=boondox,calscaled=1]{mathalfa}
\DeclareMathAlphabet{\bbvar}{U}{BOONDOX-ds}{m}{n}
\DeclareMathAlphabet{\bbgreek}{U}{bbold}{m}{n}
\usepackage{psfrag}
\usepackage{pstool}
\usepackage{caption}
\usepackage{tabularx}
\usepackage{multirow}
\usepackage{pbox}
\usepackage{graphicx}
\newcommand{\hook}{\text{\large{$\lrcorner$}}}

\usepackage[colorlinks=true,
            linkcolor=blue,
            urlcolor=blue,
            citecolor=blue]{hyperref}
\usepackage[hyperref]{xcolor}
\definecolor{darkgreen}{rgb}{0.01, 0.75, 0.24}

\newcommand{\qq}[1]{``#1''} %Anfuehrungszeichen
\newcommand{\di}{\mathrm{d}}
\usepackage{tensor}
\newcommand{\ou}[3]{\tensor{#1}{^{#2}_{#3}}}
\newcommand{\uo}[3]{\tensor{#1}{_{#2}^{#3}}}

\newcommand{\I}{\mathrm{i}} %imaginaere Einheit
\newcommand{\CC}{\mathrm{cc.}} % komplex konjugiertes
\newcommand{\C}{\mathbb{C}}

\newcommand{\R}{\mathbb{R}}

\newcommand{\eref}[1]{(\ref{#1})}

\newcommand{\bbwedge}{\reflectbox{\rotatebox[origin=c]{180}{\fontsize{10pt}{10pt}$\hspace{0.7pt}\bbvar{V}\hspace{0.7pt}$}}}

\newcommand{\mtext}[1]{\text{\it #1}}
\usepackage{tikz-cd}

\newcommand\vpm{\mathbin{\vcenter{\hbox{
  \oalign{\hfil$\scriptstyle+$\hfil\cr
          \noalign{\kern-.3ex}
          $\scriptscriptstyle({-})$\cr}}}}}
\DeclareMathAlphabet{\sfit}{OT1}{fos}{sb}{it}
\DeclareMathAlphabet{\mathsf}{OT1}{fos}{sb}{n}

\definecolor{darkgreen}{rgb}{0.01, 0.75, 0.24}
\definecolor{darkblue}{rgb}{0.01, 0.24, 0.75}

\usepackage[multiple, flushmargin]{footmisc}

\let\originalleft\left
\let\originalright\right
\renewcommand{\left}{\mathopen{}\mathclose\bgroup\originalleft}
\renewcommand{\right}{\aftergroup\egroup\originalright}

\newcommand{\dbarvar}{{\mathrm{d}\mkern-7.5mu\lower.18ex\hbox{$\textasciitilde$}\mkern-1.5mu}}

\begin{document}

\begin{abstract}
\noindent The Barnich--Troessaert bracket is a proposal for a modified Poisson bracket on the covariant phase space for general relativity. The new bracket allows us to compute charges, which are otherwise not integrable. Yet there is a catch. There is a clear prescription for how to evaluate the new bracket for any such charge, but little is known how to extend the bracket to the entire phase space. This is a problem, because not every gravitational observable is also a charge.  In this paper, we propose such an extension. The basic idea is to remove the radiative data from the covariant phase space. This requires second-class constraints. Given a few basic assumptions, we show that the resulting Dirac bracket on the constraint surface is nothing but the BT bracket. A heuristic argument is given to show that the resulting constraint surface can only contain gravitational edge modes.
%The Barnich\,--\,Troessaert (BT) bracket is a proposal for a modified Poisson bracket on the covariant phase space. It allows us to compute charges, which are otherwise not integrable. Yet there is a catch. By its very definition, the new bracket can only be evaluated between  charges. This is a problem, because phase space is vastly bigger. Not every observable is also a charge.  In this paper, we propose  an extension of the BT bracket to the entire phase space. The basic idea is to remove the radiative data from the covariant phase space. This requires second-class constraints. Given a few basic assumptions, we show that the resulting Dirac bracket is the BT bracket. The construction defines a reduced phase space, which only contains Coulombic modes of the gravitational field ( gravitational edge modes).
\end{abstract}%
\title{Barnich--Troessaert Bracket as a Dirac Bracket on the Covariant Phase Space}
\author{Wolfgang Wieland}
\address{Institute for Quantum Optics and Quantum Information (IQOQI)\\Austrian Academy of Sciences\\Boltzmanngasse 3, 1090 Vienna, Austria\\{\vspace{0.5em}\normalfont 23 November 2021}
}
%\date{Winter 2017}
%\keywords{test}
\maketitle
\vspace{-1.2em}
\hypersetup{
  linkcolor=black,
  urlcolor=black,
  citecolor=black
}
{\tableofcontents}
%:
\hypersetup{
  linkcolor=black,
  urlcolor=darkgreen,
  citecolor=darkgreen,
}
\begin{center}{\noindent\rule{\linewidth}{0.4pt}}\end{center}%\newpage
\section{Introduction}
\noindent At null infinity, there is no conserved mass, because gravitational radiation carries energy \cite{Bondi21,Sachs103,Horowitz:1981uw,AshtekarNullInfinity,Ashtekar:2014zsa}. An immediate consequence of this simple observation is that the BMS supertranslations are not integrable on the covariant phase space \cite{Peierls,Ashtekar:1990gc,Lee:1990nz,Iyer:1994ys,Wald:1999wa}. If $\Omega_M(\cdot,\cdot)$ is the (vastly degenerate) pre-symplectic two-form for a partial Cauchy surface $M$ that intersects $\mathcal{I}^+$ at a cross section $\partial M=\mathcal{C}\subset\mathcal{I}^+$, the relevant equation reads
\begin{align}\nonumber
\Omega_{M}(\mathcal{L}_\xi,\delta)=\frac{1}{8\pi G}\oint_{\partial \mathcal{C}}\xi^ak_a\Big[&\kappa_{(\ell)}\delta\varepsilon-\tfrac{1}{2}\vartheta_{(\ell)}\delta\varepsilon-\varepsilon\delta\vartheta_{(\ell)}\\
&-\I\,\sigma_{(\ell)}\bar{m}\wedge\delta\bar{m}+\I\,\bar{\sigma}_{(\ell)}m\wedge\delta m\Big]\neq-\delta[Q_\xi],\label{def1}
\end{align}
where $\xi^a$ is a BMS supertranslation, $\ell^a$ denotes a null generator of $\mathcal{J}^+$, $\kappa_{(\ell)}$ is its non-affinity, and $m_a,\bar{m}_a$ is a $U(1)$ dyad on the null surface, such that $2\,m_{(a}\bar{m}_{b)}$ is the pull-back of the spacetime metric $g_{ab}$ to the null boundary. The canonical area element is $\varepsilon=-\I\,m\wedge\bar{m}$, and $\vartheta_{(\ell)}$ and $\sigma_{(\ell)}$ are the shear and expansion of the null generator, while the one-form  $k_a$ is dual to it, i.e.\ $k_a\ell^a=-1$. Equation \eref{def1} holds for generic null surfaces \cite{Wieland:2017zkf,Wieland:2020gno, Wieland:2021vef}. Taking into account the fall-off and gauge-fixing conditions on $\mathcal{I}^+$, the  terms that are responsible for the non-integrability of $Q_\xi$ are only contained in the second line, which depends on the time derivative $\dot{\sigma}^{(0)}(u,z,\bar{z})$ of the asymptotic shear via $\sigma_{(\ell)}=-\dot{\sigma}^{(0)}(u,z,\bar{z})/r+\mathcal{O}(r^{-2})$, see e.g.\ \cite{Wieland:2020gno}.

To compute the charges from the pre-symplectic two-form, we have to relax the requirement that $Q_\xi$ is the Hamiltonian generator of the desired symmetries. This can be achieved by adding a counter term, which depends on the  symplectic current $J_{\mtext{rad}}$ of the radiative modes. This counter term was  identified by Wald and Zoupas in \cite{Wald:1999wa}. Computing the resulting charge amounts to integrating the equation
\begin{equation}
\delta[Q_\xi] = - \Omega_{M}(\mathcal{L}_\xi,\delta)+\oint_{\mathcal{C}}\xi\hook J_{\mtext{rad}}(\delta),\label{def2}
\end{equation}
for all linearised solutions $\delta[\cdot]$ on the covariant phase space. More recent results have given prescriptions to extend these definitions to finite domains, see \cite{Chandrasekaran:2018aop,Wieland:2020gno,Wieland:2021vef,Chandrasekaran:2020wwn}. Equation \eref{def2} defines a charge, but now we face the problem that we cannot use covariant phase space methods to compute the resulting commutation relations $\{Q_\xi,Q_{\xi'}\}$, because the Hamiltonian vector field of $Q_\xi$, so it exists, does not coincide with the Lie derivative $\mathcal{L}_\xi$. \medskip

A proposal to resolve this issue was given by Barnich and Troessaert, who introduced a new bracket \cite{Barnich:2009se,Barnich:2011mi,Barnich:2010eb}. On the covariant phase space, it is defined as follows: if $J_{\mtext{rad}}$ denotes  the  symplectic current for the radiative modes at null infinity, the new bracket is given by
\begin{equation}
\big\{Q_\xi,Q_{\xi'}\big\}_{\mathrm{TB}} := \Omega_M(\mathcal{L}_\xi,\mathcal{L}_{\xi'}) - \oint_{\mathcal{C}}\Big[\xi\hook J_{\mtext{rad}}(\mathcal{L}_{\xi'})-\xi'\hook J_{\mtext{rad}}(\mathcal{L}_{\xi})\Big].
\end{equation}
Now we have a new bracket, but by changing the bracket, we also change the phase space. Therefore, a new set of questions arises. What is the phase space for which the Barnich\,--\,Troessaert bracket defines a (non-degenerate) symplectic two-form? Furthermore, if $O$ and $O'$ denote Dirac observables of the gravitational field, such as those defined in e.g.\ \cite{DittTambo1,Dittrich:2005kc,Giddings2005,DonnellyGiddings2016}, what are their commutation relations with respect to the new bracket, i.e.\ what is 
$\{O,O'\}_{\mathrm{TB}}$ for generic Dirac observables $O$ and $O'$?
\medskip

 In this note, we will reflect on these questions. Our main message will be that the Barnich--Troessaert bracket should be understood as an ordinary Dirac bracket for a large (in fact infinite) number of second-class constraints. The role of the second-class constraints is to simply remove the radiative data from the covariant phase space on a partial Cauchy surface $M$ and replace them by auxiliary background fields ($c$-numbers). The resulting reduced phase space, which is now indexed by the background fields, is the phase space of gravitational edge modes alone.  A different and more algebraic perspective is given in \cite{Freidel:2021yqe}.

\begin{figure}[h]
\begin{center}
\psfrag{I}{$\mathcal{I}^+$}
\psfrag{C1}{$\mathcal{C}_+$}
\psfrag{M0}{$M$}
\psfrag{C0}{$\mathcal{C}$}
\psfrag{M1}{$M_+$}
\psfrag{B}{$\mathcal{M}$}
\psfrag{k}{$i_+$}
%\psfrag{c}{$f$}
\hspace{1.5em}\includegraphics[width=0.29\textwidth]{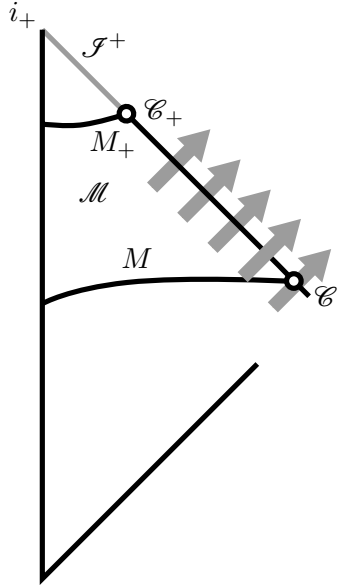}
\end{center}
\caption*{Figure 1: Setup of the problem. We consider an asymptotically flat spacetime. The three-manifolds $M$ and $M_+$ are partial Cauchy hypersurfaces, which are bounded by consecutive cross sections $\mathcal{C}=\partial M$ and $\mathcal{C}_+=\partial M_+$ of future null infinity $\mathcal{I}^+$.  The null surface $\mathcal{N}$ is the portion of $\mathcal{I}^+$ between $\mathcal{C}$ and $\mathcal{C}_+$. We restrict ourselves to regions in phase space where $\mathcal{C}_+$ lies far enough ahead such that all radiation at $\mathcal{I}^+$ vanishes at and beyond the  cross section $\mathcal{C}_+$. Care needs to be taken with orientations. Our conventions are as follows. The orientation of $\mathcal{N}$ is induced from the bulk, which is $\mathcal{M}$, whereas the orientation of the cross sections $\{\mathcal{C},\mathcal{C}_+\}$ is induced from $M$ and $M_+$. The boundary of 
$\mathcal{M}$ is $\partial\mathcal{M}=M\cup M_+^{-1}\cup\mathcal{N}$.}\label{fig1}
\end{figure}

\section{Covariant phase space and bulk-boundary dynamics}
\noindent To begin with, let us first clarify the problem in full generality. The analysis will be based on covariant phase space methods for manifolds with boundaries \cite{Peierls,Ashtekar:1990gc,Lee:1990nz,Iyer:1994ys,Wald:1999wa,Wieland:2020gno,Barbero2021qiz,Harlow:2020aa}. The general set up is a field theory on a $d$-dimensional manifold $\mathcal{M}$ with a time-like or null boundary. Typically, its topology is that of an infinite\footnote{This is in slight derogation from \hyperref[fig1]{figure 1}, where the null surface $\mathcal{N}$ has itself a boundary.} cylinder, i.e.\ $\mathcal{N}=S^{d-2}\times\R$,  $\partial\mathcal{M}=\mathcal{N}$. The bulk and boundary field equations are derived from the variation of an action, which will have the following general form
\begin{align}
S[\Phi,\varphi|\sigma] = \int_{\mathcal{M}}L[\Phi,\di\Phi] + \int_{\mathcal{N}}l[\Phi,\varphi,\di\varphi|\sigma],\label{actndef}
\end{align}
where the $\mathbb{R}$-valued $d$-form $L[\Phi,\di\Phi]\in\Omega^d(\mathcal{M}:\mathbb{R})$ is a Lagrangian in the bulk and $l[\Phi,\varphi,\di\varphi|\sigma]\in\Omega^{d-1}(\mathcal{N}:\mathbb{R})$ is the boundary Lagrangian.  The action \eref{actndef} is a functional 
\begin{equation}
S:\mathcal{F}_{\mtext{kin}}\rightarrow\mathbb{R};(\Phi,\varphi|\sigma)\mapsto S[\Phi,\varphi|\sigma],
\end{equation}
on the space of kinematical histories, i.e.\ the space of bulk and boundary field configurations for $(\Phi, \varphi|\sigma)$. The basic configuration variables are $\Phi$ and $\varphi$, which are tensor-valued\footnote{In the following, all tensor indices are suppressed, and \qq{$\di$} denotes the exterior derivative.}  differential forms, i.e.\ $\Phi\in\Omega^{|\Phi|}(\mathcal{M}:\mathbb{V})$ and $\varphi\in\Omega^{|\varphi|}(\mathcal{N}:\mathbb{W})$ that take values in  some unspecified target spaces $\mathbb{V}$ and $\mathbb{W}$. The integer $|\Phi|=p$ is the degree of the $p$-form $\Phi$. The bulk and boundary Lagrangians  depend only on the fields and their exterior derivatives. In the absence of a metric or other background structures, the only available derivative that can operate on a $p$-form and commutes with the pull-back is the exterior derivative, which is nilpotent, i.e.\ $\di^2\Phi=0$. Hence, no second derivatives can ever appear in our Lagrangian.\footnote{Second derivatives can only appear by integrating out auxiliary fields. This happens when going from the Einstein--Palatini--Cartan action to the more familiar Einstein--Hilbert action, where we solve for the torsion-free condition and insert it back into the action. } Besides the configuration variables, the action also depends on external sources $\sigma$ ($c$-numbers or background fields), which are unspecified tensor-valued $p$-forms on the boundary. Both the bulk and boundary Lagrangians are local in all fundamental variables. Examples for such bulk and boundary actions in three and four spacetime dimensions are plentiful, see e.g.\ \cite{Wieland:2018ymr,Wieland:2020gno,Freidel:2020xyx,Freidel:2020svx,Geiller:2019bti,Margalef-Bentabol:2020teu,Harlow:2020aa} and references therein.

Since we are considering gravity, the action is diffeomorphism invariant. For all $\alpha\in\mathrm{Diff}(\mathcal{M}:\mathcal{M})$, we thus have\footnote{We may assume that the bulk Lagrangian $L[\Phi,\di\Phi]$ is invariant under diffeomorphisms only up to a total exterior derivative, but such exact forms should be reabsorbed into a redefinition of the boundary Lagrangian $l$.}
\begin{align}
L[\alpha^\ast\Phi,\di(\alpha^\ast\Phi)]&=L[\alpha^\ast\Phi,\alpha^\ast(\di\Phi)]\approx(\alpha^\ast L)[\Phi,\di\Phi],\\
l[\alpha^\ast\Phi,\alpha^\ast\varphi,\di(\alpha^\ast\varphi)|\alpha^\ast\sigma]&=l[\alpha^\ast\Phi,\alpha^\ast\varphi,\alpha^\ast(\di\varphi)|\alpha^\ast\sigma]=(\alpha^\ast l)[\Phi,\varphi,\di\varphi|\sigma],%\\
\end{align}
where $\alpha^\ast$ denotes the pull-back.

To introduce the covariant Hamiltonian formalism, which equips $\mathcal{F}_{\mtext{kin}}$ with a pre-symplectic structure, it is useful to define the \emph{kinetic momenta},\footnote{The terminology is borrowed from electrodynamics. The kinetic momentum of a charged particle is its four-velocity $\dot{x}^\mu$, the canonical momentum, on the other hand is $p_\mu = m\dot{x}_\mu-e A_\mu(x)$, where $A_\mu$ is the vector potential.}
\begin{align}
\Pi_\Phi = \di\Phi\in\Omega^{|\Phi|+1}(\mathcal{M}:\mathbb{V}),\\
\pi_\varphi = \di\varphi\in\Omega^{|\varphi|+1}(\mathcal{M}:\mathbb{W}).
\end{align}
The field equations follow from the variation of the action. If $\delta\in T\mathcal{F}_{\mtext{kin}}$ denotes a tangent vector (variation) on field space, we have\footnote{N.B.: if $\Phi\in\Omega^{|\Phi|}(\mathcal{M},\mathbb{V})$ is a $\mathbb{V}$-valued differential form, the derivative $\frac{\partial L}{\partial \Phi}$ defines a $(d-p)$-form that takes values in the dual vector space $\mathbb{V}^\ast$. Accordingly, the symbol {$\mathrm{Tr}$} denotes the natural pairing between elements of $\mathbb{V}$ and $\mathbb{V}^\ast$.}
\begin{align}
 \delta[L] &= \mathrm{Tr}\Big[\frac{\partial L}{\partial\Phi}\wedge\delta[\Phi]+\frac{\partial L}{\partial\Pi_\Phi}\wedge\delta[\Pi_\Phi]\Big]=(\mathrm{EOM})(\delta)+\di[J_{\mtext{bulk}}(\delta)],\label{L-var}
\end{align}
where we defined the following one-forms on field space $\mathcal{F}_{\mtext{kin}}$, namely
\begin{align}
\mathrm{EOM} & =  \mathrm{Tr}\Big[\Big(\frac{\partial L}{\partial \Phi}+(-1)^{d-|\Phi|}\di\Big(\frac{\partial L}{\partial\Pi_\Phi}\Big)\Big)\wedge\bbvar{d}\Phi\Big],\\
J_{\mtext{bulk}} &= (-1)^{d-|\Phi|-1}\mathrm{Tr}\Big[\frac{\partial L}{\partial\Pi_\Phi}\wedge\bbvar{d}\Phi\Big],
\end{align}
and $\bbvar{d}$ denotes the exterior derivative on the space of kinematical histories $\mathcal{F}_{\mtext{kin}}$. In the same way, we introduce the variation of the boundary Lagrangian,
\begin{align}\nonumber
\delta[l]&=\mathrm{Tr}\Big[\frac{\partial l}{\partial\Phi}\wedge\delta[\Phi]+\frac{\partial l}{\partial\varphi}\delta[\varphi]+\frac{\partial l}{\partial\pi_\varphi}\delta[\pi_\varphi]+\frac{\partial l}{\partial\sigma}\delta[\sigma]\Big]=\\
&=(\mathrm{eom})(\delta)-\di[j_{\mtext{edge}}(\delta)]-J_{\mtext{glue}}(\delta)+J_{\mtext{source}}(\delta),\label{l-var}
\end{align}
where we introduced the following one-forms on field space
\begin{align}
\mathrm{eom}&=\mathrm{Tr}\Big[\Big(\frac{\partial l}{\partial\varphi}+(-1)^{d-1-|\varphi|}\di\Big(\frac{\partial l}{\partial\pi_\varphi}\Big)\Big)\wedge\bbvar{d}\varphi\Big],\\
J_{\mtext{glue}}& = -\mathrm{Tr}\Big[\frac{\partial l}{\partial\Phi}\wedge\bbvar{d}\Phi\Big],\\
J_{\mtext{source}}& = -\mathrm{Tr}\Big[\frac{\partial l}{\partial\sigma}\wedge\bbvar{d}\sigma\Big],\\
j_{\mtext{edge}}&=(-1)^{d-1-|\varphi|}\mathrm{Tr}\Big[\frac{\partial l}{\partial\pi_\varphi}\wedge\bbvar{d}\varphi\Big].
\end{align}

At its saddle points, the coupled bulk plus boundary action is stationary under all variations $\delta\in T\mathcal{F}_{\mtext{kin}}$ that satisfy the boundary conditions, which are now given by
\begin{equation}
\int_{\mathcal{N}}J_{\mtext{source}}(\delta)=0.
\end{equation}
The resulting bulk and boundary field equations are $\mathrm{EOM}=0$, $\mathrm{eom}=0$ plus additional gluing conditions. The {gluing conditions} couple the boundary fields (i.e.\ $\varphi$ and $\sigma$) to the pull-back (i.e.\ $\alpha^\ast_{\mathcal{N}}\Phi$) of the configuration variables in the bulk. The solutions to the bulk and boundary field equations and gluing conditions define the space of physical histories  $\mathcal{F}_{\mtext{phys}}\hookrightarrow\mathcal{F}_{\mtext{kin}}$, where for all $\delta\in T\mathcal{F}_{\mtext{kin}}$,
\begin{align}
(\mathrm{EOM})(\delta)\big|_{\mathcal{F}_{\mtext{phys}}}&=0,\label{bulkEOM}\\
(\mathrm{eom})(\delta)\big|_{\mathcal{F}_{\mtext{phys}}}&=0,\label{boundaryEOM}\\
(\alpha^\ast_{\mathcal{N}}J_{\mtext{bulk}})(\delta)\big|_{\mathcal{F}_{\mtext{kin}}}-J_{\mtext{glue}}(\delta)\big|_{\mathcal{F}_{\mtext{phys}}}&=0,\label{glucond}
\end{align}
where $\alpha^\ast_{\mathcal{N}}:T^\ast\mathcal{M}\rightarrow T^\ast\mathcal{N}$ denotes the pull-back of differential forms from the interior of the manifold to the boundary. 

The pre-symplectic  currents $J_{\mtext{bulk}}$ and $j_{\mtext{edge}}$ define the pre-symplectic potential, which, in turn, defines the pre-symplectic structure on the covariant phase space. Given a partial Cauchy surface $M$, which is anchored at the boundary, i.e.\ $\partial{M}=\mathcal{C}\subset\mathcal{N}$, we obtain the pre-symplectic potential

\begin{equation}
\Theta_M = \int_M J_{\mtext{bulk}}+\oint_{\partial M}j_{\mtext{edge}}.\label{Thetadef}
\end{equation}

The pre-symplectic two-form $\Omega_M$ is the exterior derivative of \eref{Thetadef}. If $\delta_1$ and $\delta_2$ are vector fields (variations) on $\mathcal{F}_{\mtext{kin}}$, and $[\delta_1,\delta_2]\in T\mathcal{F}_{\mtext{kin}}$ denotes their Lie bracket, we have
\begin{equation}
\Omega_M(\delta_1,\delta_2)=\delta_1\Big[\Theta_M(\delta_2)\Big]-\delta_2\Big[\Theta_M(\delta_1)\Big]-\Theta_M\big([\delta_1,\delta_2]\big).
\end{equation}

\section{Boundary Hamiltonian and Hamiltonian flux}
\noindent Next, we introduce a quasi-Hamiltonian on the space of physical histories. In gravity, diffeomorphisms are gauge symmetries. For every gauge symmetry, there is a corresponding conserved current, which is the exterior derivative of some charge aspect. The resulting total charge on a $(d-1)$-dimensional surface $M$ will vanish unless there is a co-dimension two boundary $\mathcal{C}=\partial M$. If there is such a boundary, the charge turns into a surface integral localised at $\mathcal{C}$. The intuitive reason why this is so is rather obvious: the introduction of the boundary breaks diffeomorphism invariance. At the boundary, there are auxiliary boundary sources $\sigma$, and the addition of these background fields breaks gauge invariance.\medskip  %What was an unphysical direction on phase space before, becomes a physical boundary mode.

At the infinitesimal level, any diffeomorphism $\alpha_\xi:\mathcal{M}\rightarrow\mathcal{M}, \alpha_\xi=\exp{\xi}$ is generated by the Lie derivative $\mathcal{L}_\xi[\cdot]=\frac{\di}{\di \varepsilon}\big|_{\varepsilon=0}\alpha_{\varepsilon\xi}^\ast$, which defines a vector field on field space, i.e.\ $\mathcal{L}_\xi[\cdot]\in T\mathcal{F}_{\mtext{kin}}$. In terms of the exterior derivative \qq{$\di$} and the interior product \qq{$\hook$},\footnote{If $\omega$ is a $p$-form, and $\xi$ is a  vector field, $(\xi\hook \omega)$ is the  $(p-1)$-form that is defined via $(\xi\hook \omega)(X_{1},\dots,X_{p-1})=\omega(\xi,X_1,\dots,X_{p-1})$ for all vector fields $X_i$.} the Lie derivative of any differential form can be written as
\begin{equation}
\mathcal{L}_\xi[\cdot] = \di\big(\xi\hook(\cdot)\big)+\xi\hook\big(\di(\cdot)\big).\label{Lxidef}
\end{equation}
In the following, we will always assume that the vector field $\xi^a\in T\mathcal{M}$ preserves the boundary, i.e.\
\begin{equation}
\xi^a\big|_{\mathcal{N}}\in T\mathcal{N},
\end{equation}
such that the definition \eref{Lxidef} naturally extends to all bulk and boundary fields $(\Phi,\varphi|\sigma)\in\mathcal{F}_{\mtext{kin}}$.  Notice also that the vector field $\xi^a\in T\mathcal{M}$ may itself be field dependent, such that e.g.\ $[\delta,\mathcal{L}_\xi]=\mathcal{L}_{\delta\xi}$ for all $\delta\in T\mathcal{F}_{\mtext{kin}}$. 

Given such a vector field $\xi^a\in T\mathcal{M}$, we may now \emph{define} the corresponding Hamiltonian as the following functional on the space of physical histories,
\begin{equation}
Q_\xi[M] := \Theta_M(\mathcal{L}_\xi)-\int_M\xi\hook L+\oint_{\partial M}\xi\hook l.\label{Hdef}
\end{equation}
Notice that we have not yet specified what the underlying phase space $\mathcal{P}$ actually is, we only used the familiar definition $H[q,\dot{q}]=\frac{\partial L}{\partial \dot{q}}\dot{q}-L[q,\dot{q}]$, where $L$ includes now both the bulk and boundary Lagrangian. Due to gauge redundancies, $\mathcal{F}_{\mtext{kin}}$ is vastly larger than $\mathcal{P}$ and it is not at all immediate to turn $H[q,\dot{q}]$ into a function on phase space (a true Hamiltonian). %We have simply defined the Hamiltonian $Q_\xi[M]$ as a functional on the space of physical histories.

The key point of this paper is to identify a candidate for a phase space, where the Hamiltonian is integrable. It is integrable if it satisfies the Hamiltonian field equations. This is to say that there is a phase space $\mathcal{P}$ and an embedding (or rather a family of gauge equivalent embeddings) $\boldsymbol{\alpha}:\mathcal{P}\hookrightarrow\mathcal{F}_{\mtext{phys}}$ such that  the pull-back $\boldsymbol{\alpha}^\ast\Omega_M$ equips $\mathcal{P}$ with a non-degenerate symplectic structure such that for all tangent vectors $\delta\in T\mathcal{P};$\begin{equation}
\delta\big[Q_\xi[M]\circ\boldsymbol{\alpha}\big] = (\boldsymbol{\alpha}_\ast\delta)\big[Q_\xi[M]\big] = \Omega_M(\boldsymbol{\alpha}_\ast\delta,\mathcal{L}_\xi)=(\boldsymbol{\alpha}^\ast\Omega_M)(\delta,\delta_\xi),
\end{equation}
where $\delta_\xi\in T\mathcal{P}$ is a tangent vector on phase space such that the difference $\boldsymbol{\alpha}_\ast\delta_\xi-\mathcal{L}_\xi$ is a null vector of the pre-symplectic potential $\Omega_M$.

To get an idea for how to construct a proposal for such an embedding, let us go again on-shell, i.e.\ restrict our discussion to $\mathcal{F}_{\mtext{phys}}$ alone. Let then $\delta$ be a linearised solution of the bulk and boundary field equations, i.e.\ a tangent vector to $\mathcal{F}_{\mtext{phys}}$.  Imposing the field equations (\ref{bulkEOM}, \ref{boundaryEOM}) and gluing conditions \eref{glucond}, and taking into account the variation of the bulk and boundary action, i.e.\ \eref{L-var} and \eref{l-var}, we obtain the variation of the Hamiltonian
\begin{align}
\nonumber\delta\big[Q_\xi[M]\big]-Q_{\delta\xi}[M]= &\delta\big[\Theta_M(\mathcal{L}_\xi)\big]-\Theta_M(\mathcal{L}_{\delta\xi})-\int_M\xi\hook\delta[L]-\oint_{\partial M}\xi\hook\delta[l]=\\
=&\delta\big[\Theta_M(\mathcal{L}_\xi)\big] -\Theta_M\big([\delta,\mathcal{L}_\xi]\big)-
\nonumber\int_M\xi\hook\di\big[J_{\mathit{bulk}}(\delta)\big]+\\
\nonumber &-\oint_{\partial M}\Big[\xi\hook\big(\di(j_{\mtext{edge}}(\delta))\big)+\xi\hook J_{\mtext{glue}}(\delta)-\xi\hook J_{\mtext{source}}(\delta)\Big]=\\
\nonumber=& \delta\big[\Theta_M(\mathcal{L}_\xi)\big] -\Theta_M\big([\delta,\mathcal{L}_\xi]\big)-
\int_M\mathcal{L}_\xi\big[J_{\mtext{bulk}}(\delta)\big]-\oint_{\partial M}\mathcal{L}_\xi\big[j_{\mtext{edge}}(\delta)\big]+\\
\nonumber&-\oint_{\partial M}\Big[-\xi\hook J_{\mtext{bulk}}(\delta)+\xi\hook J_{\mtext{glue}}(\delta)-\xi\hook J_{\mtext{source}}(\delta)\Big]=\\
=& \delta\big[\Theta_M(\mathcal{L}_\xi)\big] -\mathcal{L}_\xi\big[\Theta_M(\delta)\big] - \Theta_M\big([\delta,\mathcal{L}_\xi]\big)+\oint_{\partial M}\xi\hook J_{\mtext{source}}(\delta).\label{deltaH}
\end{align}
In other words,
\begin{equation}
\delta\big[Q_\xi[M]\big]-Q_{\delta\xi}[M]=\Omega_M(\delta,\mathcal{L}_\xi)+{\oint_{\partial M}}\xi\hook J_{\mtext{source}}(\delta).\label{charg-var}
\end{equation}
If we insist to use only field-independent diffeomorphisms, i.e.\ $\delta\xi^a =0$, the Hamiltonian is integrable only on those  surfaces in field space, where the variation of the source term ${\oint_{\partial M}}\xi\hook J_{\mtext{source}}(\delta)$ is constrained to vanish. In general, ${\oint_{\partial M}}\xi\hook J_{\mtext{source}}(\delta)\neq 0$ and the Hamiltonian is non-integrable.  In three spacetime dimensions, this is not a big deal. Choosing e.g.\ conformal boundary conditions, see e.g. \cite{Witten:2018lgb,Wieland:2018ymr}, the source $\sigma$ is essentially the conformal class $\tilde{q}^{ab}=d^2x \sqrt{q}\,q^{ab}$ of the fiducial boundary metric. In three-dimensions, the constraint $\delta[\tilde{q}^{ab}]=0$ is no restriction to the space of solutions to the field equations in the bulk, which is the moduli space of flat connections in the interior. Infinitesimal boundary diffeomorphism that preserve the background structure $\tilde{q}^{ab}$ are  generated by conformal Killing vectors $\xi^a:\mathcal{L}_\xi\tilde{q}^{ab}=0$. For every such conformal Killing $\xi^a$, there is a corresponding conserved charge (essentially a Virasoro generator). In higher dimensions, the situation is very different, because now gravity is  no longer topological. There are gravitational waves, and for generic boundaries,\footnote{If we restrict ourselves to spacelike infinity, the term ${\oint_{\partial M\rightarrow i_o}}\xi\hook J_{\mtext{source}}(\delta)=0$ will vanish for asymptotic symmetries due to the falloff and parity conditions at $i_o$.} the imposition of ${\oint_{\partial M}}\xi\hook J_{\mtext{source}}(\delta)=0$ will be a very strong constraint on the solutions in the bulk. \medskip

Let us close this section by considering the finite flux version of the infinitesimal variation \eref{deltaH} that we identified above. Consider thus two consecutive slices $M$ and $M_+$, with $\mathcal{N}_{M\rightarrow M_+}$ denoting the portion of $\mathcal{N}$ between $M$ and $M_+$, such that $M$, $M_+$ and $\mathcal{N}_{M\rightarrow M_+}$ bound a $d$-dimensional region $\mathcal{M}\subset\mathcal{M}:\partial\mathcal{M}=M^{-1}\cup M_+\cup\mathcal{N}_{M\rightarrow M_+}$, see \hyperref[fig1]{figure 1}. Going back to the variation of the bulk and boundary Lagrangian, i.e. \eref{L-var} and \eref{l-var}, and taking into account that $\xi^a\big|_{\mathcal{N}}\in T\mathcal{N}$, we obtain
\begin{align}
\nonumber Q_\xi[M_+]-Q_\xi[M]=& \int_{\partial{\mathcal{M}}^\ast}J_{\mtext{bulk}}(\mathcal{L}_\xi)-\int_{\mathrlap{\mathcal{N}_{M\rightarrow M_+}}}\quad J_{\mtext{glue}}(\mathcal{L}_\xi)-\int_{\mathrlap{\mathcal{N}_{M\rightarrow M_+}}}\quad\di\big[j_{\mtext{edge}}(\delta)\big]+\\\nonumber
&-\int_{\partial\mathcal{M}}\xi\hook L+\int_{\mathrlap{\mathcal{N}_{M\rightarrow M_+}}}\quad\xi\hook L\,+\oint_{\partial M_+}\xi\hook l-\oint_{\partial M}\xi\hook l=\\
\nonumber  =& \int_{\mathcal{M}}\mathcal{L}_\xi[L]-\int_{\mathrlap{\mathcal{N}_{M\rightarrow M_+}}}\quad J_{\mtext{glue}}(\mathcal{L}_\xi)+\int_{\mathrlap{\mathcal{N}_{M\rightarrow M_+}}}\quad\Big[\mathcal{L}_\xi[l]+J_{\mtext{glue}}(\mathcal{L}_\xi)-J_{\mtext{source}}(\mathcal{L}_\xi)\Big]+\\
\nonumber  & -\int_{\partial\mathcal{M}}\xi\hook L+\int_{\mathrlap{\mathcal{N}_{M\rightarrow M_+}}}\quad\xi\hook L\,-\int_{\mathrlap{\mathcal{N}_{M\rightarrow M_+}}}\quad\mathcal{L}_\xi[l]=\\
\nonumber  =& -\int_{\mathrlap{\mathcal{N}_{M\rightarrow M_+}}}\quad\Big[J_{\mtext{source}}(\mathcal{L}_\xi)-\xi\hook L\Big]=-\int_{\mathrlap{\mathcal{N}_{M\rightarrow M_+}}}\quad J_{\mtext{source}}(\mathcal{L}_\xi),
\end{align}
where we assumed that the gluing conditions and bulk and boundary field equations are satisfied. We thus have a balance law
\begin{equation}
F_\xi[M\rightarrow M_+]:=Q_\xi[M_+]-Q_\xi[M]=-\int_{\mathrlap{\mathcal{N}_{M\rightarrow M_+}}}\quad J_{\mtext{source}}(\mathcal{L}_\xi).\label{Hflux}
\end{equation}
If the Lie derivative $\mathcal{L}_\xi$ does not preserve the background fields (the sources $\sigma$), the Hamiltonian will  not be conserved in general. A more simplified derivation of  the flux law \eref{Hflux} and the variation of the charge \eref{charg-var} based on functional-differential techniques has been recently developed by Margalef-Bentabol and Villaseñor, see section III of \cite{Margalef-Bentabol:2020teu}. \medskip

All such derivations for charges and fluxes based on the covariant phase space approach hide an important subtlety. The existence of the BT bracket is a manifestation of this fact. The basic problem is that the covariant phase space approach is based on field space, but field space is much bigger than phase space. Phase space $\mathcal{P}$ is a submanifold $\mathcal{P}\hookrightarrow\mathcal{F}_{\mtext{phys}}\hookrightarrow\mathcal{F}_{\mtext{kin}}$, where the pre-symplectic two-form has no null directions. There is no unique such phase space, because the embedding depends on various gauge-fixing, boundary and falloff conditions. If we have found an observable $O$, which is integrable,\footnote{This is to say $\delta[O] = \Omega(\delta,\mathbb{X}_O)$ for a (Hamiltonian) vector field $\mathbb{X}_O\in T\mathcal{F}_{\mtext{phys}}$ and  all variations $\delta$ that satisfy the boundary and falloff conditions.} we are still left with the difficult task to compute the resulting Poisson brackets $\{O,O'\}_{\mathcal{P}}$. To equate  $\{\cdot,\cdot\}_{\mathcal{P}}$ with $\Omega(\mathbb{X}_O,\mathbb{X}_{O'})$ is only possible if the vector fields $\mathbb{X}_O$ ($\mathbb{X}_{O'}$) lie tangential to $\mathcal{P}$. In general $\{O,O'\}_{\mathcal{P}}\neq\Omega(\mathbb{X}_O,\mathbb{X}_{O'})$. In the following, we will see that the BT bracket provides a specific example of this subtlety on a natural phase space attached to future infinity $i_+$.
\section{Barnich--Troessaert Bracket as a Dirac Bracket}

\noindent For simplicity and definiteness, we consider here the Palatini action in asymptotically flat spacetimes. The action is evaluated in a four-dimensional spacetime region $\mathcal{M}$ that bounds future null infinity, 
\begin{equation}
S[e,A]=\frac{\I}{8\pi G}\left[\int_{\mathcal{M}}\Sigma_{AB}\wedge F^{AB}+\int_{\mathcal{N}}\eta_A\wedge\left(D-\frac{1}{2}\varkappa\right)\ell^A\right]+\CC
\end{equation}
The fields in the interior are the self-dual connection $\ou{A}{A}{B}$, whose curvature is $\ou{F}{A}{B}$, and the soldering forms $e_{AA'}$.  The soldering forms determine the two-form $\Sigma_{AB}$ (the Pleba\'{n}ski two-form), which is the self-dual part of $e_{AA'}\wedge e_{BB'}$, i.e.\ $\Sigma_{AB}=-\frac{1}{2}e _{AC'}\wedge \uo{e}{B}{C'}$. The natural $SL(2,\C)$ covariant derivative at the boundary is $D$, which is $D=\di+[\alpha^\ast_{\mathcal{N}}A,\cdot]$. The boundary fields are the null flag $\ell^A$, the spinor-valued two-form $\eta_A$, and the abelian boost connection $\varkappa$, which defines the non-affinity of the null generators, see \cite{Wieland:2021vef,Wieland:2017zkf}. 

At the saddle points, where the bulk and boundary field equations are satisfied, the variation of the action is determined by the pre-symplectic potentials $\Theta$ on the various components of the boundary. 
\begin{equation}
\delta[S]\Big|_{\mathrm{EOM}=0} = \Theta_{M}(\delta)-\Theta_{M_+}(\delta)+\Theta_\mathcal{N}(\delta),
\end{equation}
where $\delta\in T\mathcal{F}_{\mtext{phys}}$ is a linearised solution of the vacuum Einstein equations for asymptotically flat boundary conditions. On $M$ and $M_+$, the pre-symplectic potential is the integral of the symplectic current, i.e.
\begin{equation}
\Theta_M = \frac{\I}{8\pi G}\left[\int_M\Sigma_{AB}\wedge\bbvar{d}A^{AB}-\oint_{\partial{M}}\eta_A\bbvar{d}\ell^A\right]+\CC\label{thetamdef}
\end{equation}

On the asymptotic boundary $\mathcal{N}$, the situation is more subtle \cite{Wald:1999wa,AshtekarNullInfinity,Ashtekar:2018lor}. We have to impose boundary and gauge fixing conditions to remove otherwise IR divergent terms.\footnote{An example for such an IR divergence arises from the naive inclusion of conformal transformations $q_{ab}^o\rightarrow\omega^2q_{ab}^o$ of the fiducial two-metric at $\mathcal{I}^+$ into the pre-symplectic potential. The constraints at null infinity impose that $\partial_u\delta\omega=0$. Such $u$-independent terms (and their coinjugate pairs) lead to IR divergent integrals at $\mathcal{I}^+$. } Upon removing such divergencies, the radiative symplectic structure \cite{AshtekarNullInfinity} is given by
\begin{equation}
\Theta_{\mathcal{N}}=-\frac{1}{8\pi G} \int_{\mathcal{N}} \di u\wedge d^2\Omega\big(\dot{\sigma}^{(0)}\delta\bar{\sigma}^{(0)}+\CC),\label{thetadef}
\end{equation}
where $\sigma^{(0)}(u,z,\bar{z})$ is the asymptotic shear and $d^2\Omega$ is the fiducial area element at $\mathcal{I}^+$. One possibility to derive the symplectic structure \eref{thetadef} is to consider the pre-symplectic radiative structure on a finite null surface and perform an asymptotic $r\rightarrow\infty$ limit using an auxiliary double-null foliation, see \cite{Wieland:2020gno,Wieland:2021vef}. \medskip

To realise the Barnich--Troessaert bracket as a Dirac bracket, we have to say what are the relevant second-class constraints. Our proposal is that the constraints \emph{remove the entire radiative data} from the covariant phase space on a partial Cauchy surface $M$. In other words, we consider the following constraints on the radiative phase space
\begin{equation}
\forall (u,z,\bar{z})\in \mathcal{N}:\Phi_\alpha\equiv \Phi[\sigma,h](u,z,\bar{z}) = \dot{\sigma}^{(0)}(u,z,\bar{z}) - \dot{h}^{(0)}(u,z,\bar{z})\approx 0,\label{consdef}
\end{equation}
where $h^{(0)}(u,z,\bar{z})$ is a background field (a $c$-number) of compact support on $\mathcal{I}^+$ that commutes with all other phase space variables and $\alpha,\beta,\gamma,\dots$ are (De Witt) multi-indices\footnote{Summation and integration over repeated pairs of such indices is implicitly assumed, i.e.\ $\sum\!\!\!\!\!\!\!\displaystyle{\int}\,\Psi^\alpha\Phi_\alpha \equiv \Psi^\alpha\Phi_\alpha$.} and the symbol \qq{$\approx$} means that the equation is imposed as a constraint. The asymptotic shear ${\sigma}^{(0)}(u,z,\bar{z})$, or more precisely its time derivative, describes the outgoing radiation. Imposing that the constraint \eref{consdef} is satisfied amounts to constraining the outgoing radiation on a portion $\mathcal{N}$ of $\mathcal{I}^+$. The constraints \eref{consdef} are second-class. In fact, the only non-vanishing Poisson brackets among ${\sigma}^{(0)}$ and ${\bar\sigma}^{(0)}$ are given by
\begin{equation}
\big\{\sigma^{(0)}(x),\bar{\sigma}^{(0)}(y)\big\} = -4\pi G\,\varTheta(x,y)\,\delta^{(2)}(x,y),\label{Poiss1}
\end{equation}
where $\varTheta$ is the step function. Since the background fields $h^{(0)}(u,z,\bar{z})$ commute under the Poisson bracket, the Dirac matrix can be inferred immediately from \eref{Poiss1}.\medskip

The Dirac bracket defines a (vastly) degenerate  pre-symplectic structure on the covariant phase space associated to $M$. Its pull-back to the constraint hypersurface $\forall\alpha:\Phi_\alpha =0$ introduces a natural pre-symplectic structure thereon. The bracket is defined as follows, see e.g.\ \cite{HennauxTeitelboim}. First of all, we have the Dirac matrix
\begin{equation}
\Delta_{\alpha\beta} = \{\Phi_\alpha,\Phi_\beta\},
\end{equation}
 Let then $\Delta^{\alpha\beta}:\Delta^{\alpha\mu}\Delta_{\mu\beta}=\delta^\alpha_\beta$ be its inverse such that we can define the resulting Dirac bracket
\begin{equation}
\big\{A,B\big\}^\ast = \big\{A,B\big\}- \big\{A,\Phi_\alpha\big\}\Delta^{\alpha\beta}\big\{\Phi_\beta, B\big\}.\label{Diracbrack}
\end{equation}
%for functionals, Dirac observables, $A$ and $B$ on the physical phase space.

Our goal is now to develop an argument to demonstrate that the Dirac bracket \eref{Diracbrack} for the constraints \eref{consdef} returns the Barnich--Troesaert bracket provided a few basic assumptions are satisfied.\footnote{We expect that some of the assumption could be dropped or weakened. In the following, we  consider, however, only the simplest possibility.} The \emph{first assumption} is that the algebra for the BMS symmetries at $\mathcal{I}^+$ as given by the Barnich--Troesaert bracket is non-anomalous. The \emph{second assumption} is that we restrict ourselves to such vector fields $\xi^a$ that have no functional dependence on the fundamental bulk and boundary fields,\footnote{An example of a field-dependent vector field would be $\xi^a[g_{ab}]=\nabla^a(R_{bcdf}R^{bcdf})$, where $R_{abcd}$ is the Riemann curvature tensor of the spacetime metric $g_{ab}$.} i.e.\ $\delta[\xi^a]=0$, $\xi^a\in T\mathcal{M}$. The \emph{third assumption} is that the outgoing radiation at $\mathcal{I}^+$ is of compact support such that there exists a cross section $\mathcal{C}^+$ beyond which no further radiation is received (see \hyperref[fig1]{figure 1} above).  The \emph{fourth assumption} is that on-shell (pull back to $\mathcal{F}_{\mtext{phys}}$) the pre-symplectic structure on $M$ admits the block-diagonal decomposition
\begin{equation}
\Omega_M = \frac{1}{2}\Omega_{\mtext{rad}}^{\alpha\beta}[\sigma,\eta]\,\bbvar{d}\sigma_\alpha\bbwedge\bbvar{d}\sigma_\beta + \frac{1}{2}\Omega_{\mtext{edge}}^{\mu\nu}[\sigma,\eta]\,\bbvar{d}\eta_\mu\bbwedge\bbvar{d}\eta_\nu,\label{Omdef}
\end{equation}
where the first term describes the radiative data on $\mathcal{N}$, but now expressed in terms of (Dirac) observables that are evaluated on $M$ rather than $\mathcal{I}^+$, whereas the second term describes all possible boundary degrees of freedom (edge modes $\eta_\mu$) that are localised at the cross section $\mathcal{C}$. The fourth assumption implies, in other words, that there is a symplectomorphism that allows us to express the radiative modes on $M$ in terms of radiative data recorded at $\mathcal{N}$, i.e.\
\begin{equation}
\Omega_{\mtext{rad}}=\frac{1}{2}\Omega_{\mtext{rad}}^{\alpha\beta}\,\bbvar{d}\sigma_\alpha\bbwedge\bbvar{d}\sigma_\beta\simeq\Omega_{\mathcal{N}},
\end{equation}
where the symbol $\simeq$ indicates that the two phase spaces are symplectomorphic. In $2+1$ or $1+1$ dimensions, the decomposition \eref{Omdef} is trivial: gravity is topological and the only contributions to the symplectic structure are the edge modes alone \cite{Balachandran:1994up,Strominger:1997eq,Banados:1998ta,carlipbook,Carlip:2005zn,Afshar:2016wfy,Compere:2017knf,Wieland:2018ymr,Namburi:2019qja,Wieland_2020}.\medskip

Since the symplectic structure factorises into edge modes and radiative modes (our fourth assumption), the inverse of the Dirac matrix is simply given by $\Delta^{\alpha\beta}\simeq\Omega_{\mtext{rad}}^{\alpha\beta}$, where the symbol \qq{$\simeq$} stands for \emph{equality under an (possibly $\eta$-dependent) symplectomorphism}. By imposing the constraint \eref{consdef}, the  corresponding pre-symplectic two-form for the Dirac bracket \eref{Diracbrack} is then only given by the contribution from the edge modes,
\begin{equation}
\Omega_M^\ast = \Omega_{\mtext{edge}} =\frac{1}{2}\Omega_{\mtext{edge}}^{\mu\nu}[h,\eta]\,\bbvar{d}\eta_\mu\bbwedge\bbvar{d}\eta_\nu.\label{Omdef2}
\end{equation}

Let us now compute charges with respect to the Dirac bracket \eref{Diracbrack}. Since we have just identified the corresponding pre-symplectic two-form, we can immediately employ covariant phase space methods to evaluate the charge (provided our assumptions are satisfied). Consider thus a tangent vector $\delta$ to the radiative phase space, such that $[\delta,\mathcal{L}_\xi]=0$. We now immediately get
\begin{align}\nonumber
\Omega_{\mtext{edge}}(\delta,\mathcal{L}_\xi) & = \Omega_M(\delta,\mathcal{L}_\xi) - \Omega_{\mtext{rad}}(\delta,\mathcal{L}_\xi)=\\\nonumber
& = \Omega_M(\delta,\mathcal{L}_\xi) - \int_{\mathcal{N}}\Big[\delta[J_{\mtext{rad}}(\mathcal{L}_\xi)]-\mathcal{L}_\xi[J_{\mtext{rad}}(\delta)]\Big]=\\
&=\Omega_M(\delta,\mathcal{L}_\xi)+\oint_{\mathcal{C}}\xi\hook J_{\mtext{rad}}(\delta) - \int_{\mathcal{N}}\delta[J_{\mtext{rad}}(\mathcal{L}_\xi)].\label{dQ1}
\end{align}
Notice that the second term is precisely the counter term, which is added in the Wald--Zoupas framework to render the pseudo-charge $\bbvar{Q}_\xi:=\Omega_M(\cdot,\mathcal{L}_\xi)$ integrable. In other words, there is a functional $Q_\xi$ on covariant phase space such that 
\begin{equation}
\delta\big[Q_\xi[\mathcal{C}]\big] = \Omega_M(\delta,\mathcal{L}_\xi)+\oint_{\mathcal{C}}\xi\hook J_{\mtext{rad}}(\delta).\label{dQ2}
\end{equation}
It is  important to note that in integrating the charges via \eref{dQ2}, the vector fields $\delta\in T\mathcal{F}_{\mtext{phys}}$  denote an arbitrary linearised solution of the field equations. It is not assumed, in particular, that they lie tangential to the constraint hypersurface \eref{consdef}, see also \cite{Wieland:2020gno}.\medskip

Thus, the first two terms of equation \eref{dQ1} reproduce the differential of the quasi-local charge \eref{charg-var}. The third term on the right hand side of \eref{dQ1} has an immediate interpretation as well, see equation \eref{Hflux} above. It determines the radiative flux associated to the asymptotic symmetry $\xi^a\big|_{\mathcal{N}}\in T\mathcal{N}$, which we assumed to be a BMS generator. Such a  flux integral can be expressed entirely in terms of radiative modes. Can it be written as the difference of two Hamiltonian generators corresponding to the two consecutive cross-sections? On the radiative phase space, this is impossible \cite{AshtekarNullInfinity}. From the perspective of the partial Cauchy hypersurface $M$, the situation is different. Now, there is a charge, and the flux is simply the difference of the charges at the two consecutive cross sections. In other words,
\begin{equation}
F_\xi[M\rightarrow M_+] = -\int_{\mathcal{N}}J_{\mtext{rad}}(\mathcal{L}_\xi) = Q_\xi[\mathcal{C}_+]-Q_\xi[\mathcal{C}]\label{Fdef}
\end{equation}
Going back to \eref{dQ1}, we obtain
\begin{equation}
\Omega_{\mtext{edge}}(\delta,\mathcal{L}_\xi) = \delta\big[Q_\xi[\mathcal{C}_+]\big]=:\delta[Q_\xi^+].\label{dQ3}
\end{equation}

Finally, let us compute the Poisson algebra for the diffeomorphism charges under the Dirac bracket. If our assumptions are satisfied, the (vastly degenerate) pre-symplectic two-form for the Dirac bracket is given by $\Omega_{\mtext{edge}}=\Omega_M-\Omega_{\mathcal{N}}\simeq \Omega_M-\Omega_{\mtext{rad}}$. Consider then vector fields $\xi^a, {\xi'}^a$, whose restriction to $\mathcal{I}^+$ is an asymptotic BMS symmetry. We now have
\begin{align}
\big\{Q_\xi^+,Q_{\xi'}^+\big\}^\ast & = \Omega_{\mtext{edge}}(\mathcal{L}_\xi,\mathcal{L}_{\xi'})  = \Omega_M(\mathcal{L}_\xi,\mathcal{L}_{\xi'}) -
\Omega_{\mtext{rad}}(\mathcal{L}_\xi,\mathcal{L}_{\xi'})= \nonumber\\
& = \Omega_M(\mathcal{L}_\xi,\mathcal{L}_{\xi'}) - \int_{\mathcal{N}}\Big[\mathcal{L}_\xi[J_{\mtext{rad}}(\mathcal{L}_{\xi'})]-\mathcal{L}_{\xi'}[J_{\mtext{rad}}(\mathcal{L}_{\xi})]-J_{\mtext{rad}}([\xi,\xi'])\Big]=\nonumber\\
& = \Omega_M(\mathcal{L}_\xi,\mathcal{L}_{\xi'}) - \oint_{\mathcal{C}}\Big[\xi\hook J_{\mtext{rad}}(\mathcal{L}_{\xi'})-\xi'\hook J_{\mtext{rad}}(\mathcal{L}_{\xi})\Big]-F_{[\xi,\xi']}[\mathcal{N}].\label{dQ4}
\end{align}
Going from the second to the third line, we inserted the definition of the radiative flux \eref{Fdef} and used Stokes' theorem and the definition of the Lie derivative $\mathcal{L}_\xi[\cdot] =\di[\xi\hook\cdot]+\xi\hook(\di[\cdot])$ to express the second and third term as an integral over the corner. There is no contribution from $\mathcal{C}_+$, because we have assumed that there is no gravitational radiation (at or) beyond $\mathcal{C}_+$.  Notice that integrals of the form $\int_{\mathcal{N}}{\xi}\hook\varphi=0$ vanish, if $\xi\in T\mathcal{N}$ and $\varphi\in \Omega^3(\mathcal{N})$.  The meaning of equation \eref{dQ4} is immediate: the first three terms are nothing but the Barnich--Troessaert bracket at the cross section $\mathcal{C}$. If the resulting algebra for the diffeomorphism charges has no anomaly (our first assumption), we obtain
\begin{align}
\Omega_M(\mathcal{L}_\xi,\mathcal{L}_{\xi'}) - \oint_{\mathcal{C}}\Big[\xi\hook J_{\mtext{rad}}(\mathcal{L}_{\xi'})-\xi'\hook J_{\mtext{rad}}(\mathcal{L}_{\xi})\Big] =-Q_{[\xi,\xi']}[\mathcal{C}]\label{TBbrack}
\end{align}
Let us now  return back to equation \eref{dQ4}. It differs from the Barnich--Troessaert bracket \eref{TBbrack} by the flux integral that simply shifts the charges upwards along the null generators. Going back to \eref{Fdef}, we obtain
\begin{align}
\big\{Q_\xi^+,Q_{\xi'}^+\big\}^\ast &=-Q_{[\xi,\xi']}[\mathcal{C}] - F_{[\xi,\xi']}[M\rightarrow M_+]= -Q_{[\xi,\xi']}^+.\label{alg}
\end{align}

We have thus given a simple argument to demonstrate that the charges $Q_\xi^+$ are integrable, but only on a reduced phase space, which is stripped off from all the radiative modes. The resulting charges satisfy the commutation relation \eref{alg}. The corresponding Poisson bracket $\{\cdot,\cdot\}^\ast$ is nothing but the Dirac bracket on the covariant phase space. The constraints \eref{consdef} remove the radiative data from the covariant phase space and turn them into auxiliary background fields on $\mathcal{C}$. Notice also that the constraints $\eref{consdef}$ will necessary commute under the Dirac bracket, since the flux only depends on the radiative modes and will thus commute under the Dirac bracket, i.e.\ $\{F_\xi,\cdot\}^\ast=0$.
\section{Summary and Conclusion}
\noindent %The radiative phase space describes the two modes of gravitational radiation at future (past) null infinity. 
On the radiative phase space, it is straightforward to introduce Hamiltonian generators for asymptotic BMS symmetries \cite{AshtekarNullInfinity,Ashtekar:2018lor}. These generators are flux integrals. They determine the evolution of the BMS charge aspect due to gravitational radiation. Yet, the charge integrals themselves do not exist on the radiative phase space. To access the charges, we need a different phase space, such as the ADM phase space \cite{adm}, which is associated to a complete Cauchy surface.  

In this note, we pointed out that there is yet another (and perhaps more minimalistic) possibility to realise the charges as Hamiltonian generators. We considered the covariant phase space on a partial Cauchy surface $M$ and removed the radiative data via the Dirac bracket. We argued that the resulting reduced phase space is the phase space of gravitational edge modes (Coulombic modes) alone. Given a few basic assumptions, we gave a heuristic argument, which allowed us to infer the resulting Dirac bracket. The result returned the Barnich--Troessaert bracket on a cross section of $\mathcal{I}^+$ plus an additional flux integral, which only depends on the radiative data, which commutes under the Dirac bracket (the flux depends only on the radiative modes). The role of the flux integral is to simply shift the charges upwards to future infinity ($i^+$). There are thus three distinct phase spaces. First of all, there is the ADM phase space on a complete Cauchy surface \cite{ADMmass}. Next, there is the radiative phase space at $\mathcal{I}^+$, which is slightly smaller. The difference between the two is also a phase space, which is the phase space of the edges modes alone, now localized at $i^+$. The symplectic structure on $i^+$ can be inferred in two different ways: via the Barnich--Troessaert bracket shifted by an additional flux integral, or via the Dirac bracket \eref{Diracbrack}.\medskip

To summarise, there are two distinct ways to consider null infinity from a Hamiltonian perspective. The first approach is to work on the usual radiative phase space, where we know the symplectic structure of the radiative data at the full non-perturbative level. On the radiative phase space, the BMS fluxes are Hamiltonian, but the charges are not. The second approach addresses this issue using a more holographic perspective. The radiative data is fixed via auxiliary boundary conditions. Imposing these boundary conditions amounts to introducing auxiliary second-class constraints such that the charges are integrable.   These background fields are not part of the resulting phase space and commute under the Dirac bracket. The holographic viewpoint clearly resonates with results in lower dimensions, where there are no radiative modes to begin with, and the entire physical phase space consist of the edge modes alone. It is our opinion that both approaches are equally important, and simply represent different ways of splitting the ADM phase space into different Hamiltonian subsystems. The question for how to identify such subsystems is an important problem both from the perspective of holography as well as non-perturbative quantum gravity and quantum foundations \cite{Andrade:2015fna,Giddings:2019hjc,Freidel:2020xyx,Freidel:2020svx,Barbero2021qiz,Donnelly:2018nbv,Wieland:2017zkf,Wieland:2017cmf,Wieland:2021vef,Wieland:2020gno,Rovelli:2020mpk,Dittrich:2018xuk,Dittrich:2017hnl,Dittrich:2017rvb,Gomes:2019xto,Harlow:2020aa,Wieland:2018ymr,Vanrietvelde:2018pgb,Hoehn:2019owq,Castro-Ruiz:2019nnl,Giacomini:2019fvi,Giacomini:2019aa,Krumm:2020fws}.

\paragraph{- Acknowledgments} The author would like to take this opportunity to thank Laurent Freidel and Simone Speziale for a fruitful email exchange. The author would also like to thank Abhay Ashtekar for very helpful comments during an ILQGS seminar in spring 2021. The author would also like to thank the reviewers, whose comments provided additional background material and helped to improve the paper.  Support from the Institute for Quantum Optics and Quantum Information is gratefully acknowledged.  This research was supported in part by the ID 61466 grant from the John Templeton Foundation, as part of The Quantum Information Structure of Spacetime (QISS) Project (qiss.fr). The opinions expressed in this publication are those of the author and do not necessarily reflect the views of the John Templeton Foundation.

%\section*{References}

\providecommand{\href}[2]{#2}\begingroup\raggedright\endgroup


\begin{thebibliography}{10}

\bibitem{Bondi21}
H.~Bondi, M.~G.~J. van~der Burg, and A.~W.~K. Metzner, ``Gravitational waves in
  general relativity, VII. Waves from axi-symmetric isolated system,'' {\em
  Proc. of the Royal Soc. Lond. A: Mathematical, Physical and Engineering
  Sciences} {\bf 269} (1962), no.~1336, 21--52.

\bibitem{Sachs103}
R.~K. Sachs, ``Gravitational waves in general relativity VIII. Waves in
  asymptotically flat space-time,'' {\em Proceedings of the Royal Society
  London A} {\bf 270} (1962), no.~1340, 103--126.

\bibitem{Horowitz:1981uw}
G.~T. Horowitz and M.~J. Perry, ``{Gravitational Energy Cannot Become
  Negative},'' {\em Phys. Rev. Lett.} {\bf 48} (1982) 371.

\bibitem{AshtekarNullInfinity}
A.~Ashtekar, {\em {Asymptotic Quantization}}.
\newblock Bibliopolis, Napoli, 1987.
\newblock Based on 1984 Naples Lectures.

\bibitem{Ashtekar:2014zsa}
A.~Ashtekar, ``{Geometry and Physics of Null Infinity},'' in {\em {Surveys in
  Differential Geometry\,---\,One hundred years of general relativity}},
  L.~Bieri and S.-T. Yau, eds., vol.~20.
\newblock International Press of Boston, 2015.
\newblock
\href{http://arXiv.org/abs/1409.1800}{{\tt arXiv:1409.1800}}.
\newblock
%%CITATION = ARXIV:1409.1800;%%.

\bibitem{Peierls}
R.~E. Peierls, ``The commutation laws of relativistic field theory,'' {\em
  Proceedings of the Royal Society of London. Series A. Mathematical and
  Physical Sciences} {\bf 214} (1952), no.~1117, 143--157,
  \href{http://arXiv.org/abs/https://royalsocietypublishing.org/doi/pdf/10.1098/rspa.1952.0158}{{\tt
  arXiv:https://royalsocietypublishing.org/doi/pdf/10.1098/rspa.1952.0158}}.

\bibitem{Ashtekar:1990gc}
A.~Ashtekar, L.~Bombelli, and O.~Reula, ``{The Covariant Phase Space Of
  Asymptotically Flat Gravitational Fields},'' in {\em Mechanics, Analysis and
  Geometry: 200 Years after Lagrange}, M.~Francaviglia and D.~Holm, eds.
\newblock North Holland, Amsterdam,
1990.
\newblock
%%CITATION = PRINT-90-0318 (SYRACUSE);%%.

\bibitem{Lee:1990nz}
J.~Lee and R.~M. Wald, ``{Local symmetries and constraints},'' {\em J. Math.
  Phys.} {\bf 31} (1990) 725--743.

\bibitem{Iyer:1994ys}
V.~Iyer and R.~M. Wald, ``{Some properties of Noether charge and a proposal for
  dynamical black hole entropy},'' {\em Phys. Rev.} {\bf D50} (1994) 846--864,
\href{http://arXiv.org/abs/gr-qc/9403028}{{\tt arXiv:gr-qc/9403028}}.
%%CITATION = GR-QC/9403028;%%.

\bibitem{Wald:1999wa}
R.~M. Wald and A.~Zoupas, ``{A General definition of `conserved quantities' in
  general relativity and other theories of gravity},'' {\em Phys. Rev. D} {\bf
  61} (2000) 084027,
\href{http://arXiv.org/abs/gr-qc/9911095}{{\tt arXiv:gr-qc/9911095}}.
%%CITATION = GR-QC/9911095;%%.

\bibitem{Wieland:2017zkf}
W.~Wieland, ``{New boundary variables for classical and quantum gravity on a
  null surface},'' {\em Class. Quantum Grav.} {\bf 34} (2017) 215008,
\href{http://arXiv.org/abs/1704.07391}{{\tt arXiv:1704.07391}}.
%%CITATION = ARXIV:1704.07391;%%.

\bibitem{Wieland:2020gno}
W.~Wieland, ``{Null infinity as an open Hamiltonian system},'' {\em JHEP} {\bf
  21} (2020) 095, \href{http://arXiv.org/abs/2012.01889}{{\tt
  arXiv:2012.01889}}.

\bibitem{Wieland:2021vef}
W.~Wieland, ``{Gravitational SL(2, \ensuremath{\mathbb{R}}) algebra on the
  light cone},'' {\em JHEP} {\bf 07} (2021) 057,
  \href{http://arXiv.org/abs/2104.05803}{{\tt arXiv:2104.05803}}.

\bibitem{Chandrasekaran:2018aop}
V.~Chandrasekaran, {\'E}.~{\'E}. Flanagan, and K.~Prabhu, ``{Symmetries and
  charges of general relativity at null boundaries},'' {\em JHEP} {\bf 11}
  (2018) 125,
\href{http://arXiv.org/abs/1807.11499}{{\tt arXiv:1807.11499}}.
%%CITATION = ARXIV:1807.11499;%%.

\bibitem{Chandrasekaran:2020wwn}
V.~Chandrasekaran and A.~J. Speranza, ``{Anomalies in gravitational charge
  algebras of null boundaries and black hole entropy},'' {\em JHEP} {\bf 01}
  (2021) 137, \href{http://arXiv.org/abs/2009.10739}{{\tt arXiv:2009.10739}}.

\bibitem{Barnich:2009se}
G.~Barnich and C.~Troessaert, ``{Symmetries of asymptotically flat 4
  dimensional spacetimes at null infinity revisited},'' {\em Phys. Rev. Lett.}
  {\bf 105} (2010) 111103, \href{http://arXiv.org/abs/0909.2617}{{\tt
  arXiv:0909.2617}}.

\bibitem{Barnich:2011mi}
G.~Barnich and C.~Troessaert, ``{BMS charge algebra},'' {\em JHEP} {\bf 12}
  (2011) 105,
\href{http://arXiv.org/abs/1106.0213}{{\tt arXiv:1106.0213}}.
%%CITATION = ARXIV:1106.0213;%%.

\bibitem{Barnich:2010eb}
G.~Barnich and C.~Troessaert, ``{Aspects of the BMS/CFT correspondence},'' {\em
  JHEP} {\bf 05} (2010) 062, \href{http://arXiv.org/abs/1001.1541}{{\tt
  arXiv:1001.1541}}.

\bibitem{DittTambo1}
B.~Dittrich and J.~Tambornino, ``{A Perturbative approach to Dirac observables
  and their space-time algebra},'' {\em Class. Quant. Grav.} {\bf 24} (2007)
  757--784,
\href{http://arXiv.org/abs/gr-qc/0610060}{{\tt arXiv:gr-qc/0610060}}.
%%CITATION = GR-QC/0610060;%%.

\bibitem{Dittrich:2005kc}
B.~Dittrich, ``{Partial and complete observables for canonical general
  relativity},'' {\em Class. Quant. Grav.} {\bf 23} (2006) 6155--6184,
\href{http://arXiv.org/abs/gr-qc/0507106}{{\tt arXiv:gr-qc/0507106}}.
%%CITATION = GR-QC/0507106;%%.

\bibitem{Giddings2005}
S.~B. Giddings, D.~Marolf, and J.~B. Hartle, ``{Observables in effective
  gravity},'' {\em Phys. Rev.} {\bf D74} (2006) 064018,
\href{http://arXiv.org/abs/hep-th/0512200}{{\tt arXiv:hep-th/0512200}}.
%%CITATION = HEP-TH/0512200;%%.

\bibitem{DonnellyGiddings2016}
W.~Donnelly and S.~B. Giddings, ``{Observables, gravitational dressing, and
  obstructions to locality and subsystems},''
\href{http://arXiv.org/abs/1607.01025}{{\tt arXiv:1607.01025}}.
%%CITATION = ARXIV:1607.01025;%%.

\bibitem{Freidel:2021yqe}
L.~Freidel, R.~Oliveri, D.~Pranzetti, and S.~Speziale, ``{The Weyl BMS group
  and Einstein's equations},'' \href{http://arXiv.org/abs/2104.05793}{{\tt
  arXiv:2104.05793}}.

\bibitem{Barbero2021qiz}
J.~F. Barbero~G., J.~Margalef-Bentabol, V.~Varo, and E.~J.~S. Villase\~nor,
  ``{Covariant phase space for gravity with boundaries: metric vs tetrad
  formulations},'' \href{http://arXiv.org/abs/2103.06362}{{\tt
  arXiv:2103.06362}}.

\bibitem{Harlow:2020aa}
D.~Harlow and J.-q. Wu, ``Covariant phase space with boundaries,'' {\em Journal
  of High Energy Physics} {\bf 2020} (2020), no.~10, 146.

\bibitem{Wieland:2018ymr}
W.~Wieland, ``{Conformal boundary conditions, loop gravity and the
  continuum},'' {\em JHEP} {\bf 10} (2018) 089,
\href{http://arXiv.org/abs/1804.08643}{{\tt arXiv:1804.08643}}.
%%CITATION = ARXIV:1804.08643;%%.

\bibitem{Freidel:2020xyx}
L.~Freidel, M.~Geiller, and D.~Pranzetti, ``{Edge modes of gravity. Part I.
  Corner potentials and charges},'' {\em JHEP} {\bf 11} (2020) 026,
  \href{http://arXiv.org/abs/2006.12527}{{\tt arXiv:2006.12527}}.

\bibitem{Freidel:2020svx}
L.~Freidel, M.~Geiller, and D.~Pranzetti, ``{Edge modes of gravity. Part II.
  Corner metric and Lorentz charges},'' {\em JHEP} {\bf 11} (2020) 027,
  \href{http://arXiv.org/abs/2007.03563}{{\tt arXiv:2007.03563}}.

\bibitem{Geiller:2019bti}
M.~Geiller and P.~Jai-akson, ``{Extended actions, dynamics of edge modes, and
  entanglement entropy},'' {\em JHEP} {\bf 09} (2020) 134,
  \href{http://arXiv.org/abs/1912.06025}{{\tt arXiv:1912.06025}}.

\bibitem{Margalef-Bentabol:2020teu}
J.~Margalef-Bentabol and E.~J.~S. Villase\~nor, ``{Geometric formulation of the
  Covariant Phase Space methods with boundaries},'' {\em Phys. Rev. D} {\bf
  103} (2021), no.~2, 025011, \href{http://arXiv.org/abs/2008.01842}{{\tt
  arXiv:2008.01842}}.

\bibitem{Witten:2018lgb}
E.~Witten, ``{A Note On Boundary Conditions In Euclidean Gravity},''
\href{http://arXiv.org/abs/1805.11559}{{\tt arXiv:1805.11559}}.
%%CITATION = ARXIV:1805.11559;%%.

\bibitem{Ashtekar:2018lor}
A.~Ashtekar, M.~Campiglia, and A.~Laddha, ``{Null infinity, the BMS group and
  infrared issues},'' {\em Gen. Rel. Grav.} {\bf 50} (2018), no.~11, 140--163,
\href{http://arXiv.org/abs/1808.07093}{{\tt arXiv:1808.07093}}.
%%CITATION = ARXIV:1808.07093;%%.

\bibitem{HennauxTeitelboim}
M.~Henneaux and C.~Teitelboim, {\em Quantization of Gauge Systems}.
\newblock Princeton University Press, Princeton, 1992.

\bibitem{Balachandran:1994up}
A.~P. Balachandran, L.~Chandar, and A.~Momen, ``{Edge states in gravity and
  black hole physics},'' {\em Nucl. Phys.} {\bf B461} (1996) 581--596,
\href{http://arXiv.org/abs/gr-qc/9412019}{{\tt arXiv:gr-qc/9412019}}.
%%CITATION = GR-QC/9412019;%%.

\bibitem{Strominger:1997eq}
A.~Strominger, ``{Black hole entropy from near horizon microstates},'' {\em
  JHEP} {\bf 02} (1998) 009,
\href{http://arXiv.org/abs/hep-th/9712251}{{\tt arXiv:hep-th/9712251}}.
%%CITATION = HEP-TH/9712251;%%.

\bibitem{Banados:1998ta}
M.~Banados, T.~Brotz, and M.~E. Ortiz, ``{Boundary dynamics and the statistical
  mechanics of the (2+1)-dimensional black hole},'' {\em Nucl. Phys. B} {\bf
  545} (1999) 340--370, \href{http://arXiv.org/abs/hep-th/9802076}{{\tt
  arXiv:hep-th/9802076}}.

\bibitem{carlipbook}
S.~Carlip, {\em Quantum Gravity in 2+1 Dimensions}.
\newblock Cambridge University Press, Cambridge, 2003.

\bibitem{Carlip:2005zn}
S.~Carlip, ``{Conformal field theory, (2+1)-dimensional gravity, and the BTZ
  black hole},'' {\em Class. Quant. Grav.} {\bf 22} (2005) R85--R124,
\href{http://arXiv.org/abs/gr-qc/0503022}{{\tt arXiv:gr-qc/0503022}}.
%%CITATION = GR-QC/0503022;%%.

\bibitem{Afshar:2016wfy}
H.~Afshar, S.~Detournay, D.~Grumiller, W.~Merbis, A.~Perez, D.~Tempo, and
  R.~Troncoso, ``{Soft Heisenberg hair on black holes in three dimensions},''
  {\em Phys. Rev. D} {\bf 93} (2016), no.~10, 101503,
\href{http://arXiv.org/abs/1603.04824}{{\tt arXiv:1603.04824}}.
%%CITATION = ARXIV:1603.04824;%%.

\bibitem{Compere:2017knf}
G.~Comp\`ere and A.~Fiorucci, ``{Asymptotically flat spacetimes with BMS$_3$
  symmetry},'' {\em Class. Quant. Grav.} {\bf 34} (2017), no.~20, 204002,
  \href{http://arXiv.org/abs/1705.06217}{{\tt arXiv:1705.06217}}.

\bibitem{Namburi:2019qja}
J.~C. Namburi and W.~Wieland, ``{Deformed Heisenberg charges in
  three-dimensional gravity},'' {\em JHEP} {\bf 03} (2020) 175,
  \href{http://arXiv.org/abs/1912.09514}{{\tt arXiv:1912.09514}}.

\bibitem{Wieland_2020}
W.~Wieland, ``Twistor representation of Jackiw--Teitelboim gravity,'' {\em
  Classical and Quantum Gravity} {\bf 37} (2020), no.~19, 195008.

\bibitem{adm}
R.~Arnowitt, S.~Deser, and C.~Misner, {\em {The dynamics of general
  relativity}}, ch.~7, pp.~227--264.
\newblock Wiley, New York, 1962.
\newblock \href{http://arXiv.org/abs/gr-qc/0405109v1}{{\tt
  arXiv:gr-qc/0405109v1}}.

\bibitem{ADMmass}
R.~Arnowitt, S.~Deser, and C.~W. Misner, ``Coordinate Invariance and Energy
  Expressions in General Relativity,'' {\em Phys. Rev.} {\bf 122} (1961)
  997--1006.

\bibitem{Andrade:2015fna}
T.~Andrade and D.~Marolf, ``{Asymptotic Symmetries from finite boxes},'' {\em
  Class. Quant. Grav.} {\bf 33} (2016), no.~1, 015013,
\href{http://arXiv.org/abs/1508.02515}{{\tt arXiv:1508.02515}}.
%%CITATION = ARXIV:1508.02515;%%.

\bibitem{Giddings:2019hjc}
S.~B. Giddings, ``{Gravitational dressing, soft charges, and perturbative
  gravitational splitting},'' {\em Phys. Rev. D} {\bf 100} (2019), no.~12,
  126001, \href{http://arXiv.org/abs/1903.06160}{{\tt arXiv:1903.06160}}.

\bibitem{Donnelly:2018nbv}
W.~Donnelly and S.~B. Giddings, ``{Gravitational splitting at first order:
  Quantum information localization in gravity},'' {\em Phys. Rev. D} {\bf 98}
  (2018), no.~8, 086006, \href{http://arXiv.org/abs/1805.11095}{{\tt
  arXiv:1805.11095}}.

\bibitem{Wieland:2017cmf}
W.~Wieland, ``{Fock representation of gravitational boundary modes and the
  discreteness of the area spectrum},'' {\em Ann. Henri Poincar{\'e}} {\bf 18}
  (2017) 3695--3717,
\href{http://arXiv.org/abs/1706.00479}{{\tt arXiv:1706.00479}}.
%%CITATION = ARXIV:1706.00479;%%.

\bibitem{Rovelli:2020mpk}
C.~Rovelli, ``{Gauge Is More Than Mathematical Redundancy},'' {\em Fundam.
  Theor. Phys.} {\bf 199} (2020) 107--110,
  \href{http://arXiv.org/abs/2009.10362}{{\tt arXiv:2009.10362}}.

\bibitem{Dittrich:2018xuk}
B.~Dittrich, C.~Goeller, E.~R. Livine, and A.~Riello, ``{Quasi-local
  holographic dualities in non-perturbative 3d quantum gravity},'' {\em Class.
  Quant. Grav.} {\bf 35} (2018), no.~13, 13LT01,
\href{http://arXiv.org/abs/1803.02759}{{\tt arXiv:1803.02759}}.
%%CITATION = ARXIV:1803.02759;%%.

\bibitem{Dittrich:2017hnl}
B.~Dittrich, C.~Goeller, E.~Livine, and A.~Riello, ``{Quasi-local holographic
  dualities in non-perturbative 3d quantum gravity I -- Convergence of multiple
  approaches and examples of Ponzano--Regge statistical duals},'' {\em Nucl.
  Phys. B} {\bf 938} (2019) 807--877,
\href{http://arXiv.org/abs/1710.04202}{{\tt arXiv:1710.04202}}.
%%CITATION = ARXIV:1710.04202;%%.

\bibitem{Dittrich:2017rvb}
B.~Dittrich, C.~Goeller, E.~R. Livine, and A.~Riello, ``{Quasi-local
  holographic dualities in non-perturbative 3d quantum gravity II -- From
  coherent quantum boundaries to BMS$_3$ characters},'' {\em Nucl. Phys. B}
  {\bf 938} (2019) 878--934,
\href{http://arXiv.org/abs/1710.04237}{{\tt arXiv:1710.04237}}.
%%CITATION = ARXIV:1710.04237;%%.

\bibitem{Gomes:2019xto}
H.~Gomes and A.~Riello, ``{The quasilocal degrees of freedom of Yang-Mills
  theory},''
\href{http://arXiv.org/abs/1910.04222}{{\tt arXiv:1910.04222}}.
%%CITATION = ARXIV:1910.04222;%%.

\bibitem{Vanrietvelde:2018pgb}
A.~Vanrietvelde, P.~A. Hoehn, F.~Giacomini, and E.~Castro-Ruiz, ``{A change of
  perspective: switching quantum reference frames via a perspective-neutral
  framework},'' {\em Quantum} {\bf 4} (2020) 225,
  \href{http://arXiv.org/abs/1809.00556}{{\tt arXiv:1809.00556}}.

\bibitem{Hoehn:2019owq}
P.~A. H\"ohn, A.~R. Smith, and M.~P. Lock, ``{The Trinity of Relational Quantum
  Dynamics},'' \href{http://arXiv.org/abs/1912.00033}{{\tt arXiv:1912.00033}}.

\bibitem{Castro-Ruiz:2019nnl}
E.~Castro-Ruiz, F.~Giacomini, A.~Belenchia, and v.~Brukner, ``{Quantum clocks
  and the temporal localisability of events in the presence of gravitating
  quantum systems},'' {\em Nature Commun.} {\bf 11} (2020), no.~1, 2672,
  \href{http://arXiv.org/abs/1908.10165}{{\tt arXiv:1908.10165}}.

\bibitem{Giacomini:2019fvi}
F.~Giacomini, E.~Castro-Ruiz, and {\v C}.~Brukner, ``{Relativistic Quantum
  Reference Frames: The Operational Meaning of Spin},'' {\em Phys. Rev. Lett.}
  {\bf 123} (2019), no.~9, 090404, \href{http://arXiv.org/abs/1811.08228}{{\tt
  arXiv:1811.08228}}.

\bibitem{Giacomini:2019aa}
F.~Giacomini, E.~Castro-Ruiz, and {\v C}.~Brukner, ``Quantum mechanics and the
  covariance of physical laws in quantum reference frames,'' {\em Nature
  Communications} {\bf 10} (2019), no.~1, 494.

\bibitem{Krumm:2020fws}
M.~Krumm, P.~A. Hoehn, and M.~P. Mueller, ``{Quantum reference frame
  transformations as symmetries and the paradox of the third particle},'' {\em
  Quantum} {\bf 5} (2021) 530, \href{http://arXiv.org/abs/2011.01951}{{\tt
  arXiv:2011.01951}}.

\end{thebibliography}
\end{document}